\documentstyle[prb,aps,twocolumn]{revtex}
\begin{document}
\draft
\flushbottom
\twocolumn[\hsize\textwidth\columnwidth\hsize\csname 
@twocolumnfalse\endcsname

\title{\bf Density functional calculations of planar DNA base-pairs}

\author{Maider Machado}

\address{
Departamento de F{\'{\i}}sica de Materiales,
Facultad de Qu{\'{\i}}mica, Universidad del Pa{\'{\i}}s Vasco,
20080 San Sebasti\'an, \\Basque Country, Spain}

\author{Pablo Ordej\'on}
\address{
Institut de Ci\`encia de Materials de Barcelona, CSIC, 08193 Bellaterra,
Barcelona, Spain}

\author{Emilio Artacho, Daniel S\'anchez-Portal
and Jos\'e M. Soler}

\address{
Instituto de Ciencia de Materiales Nicol\'as Cabrera and
Departamento de F{\'{\i}}sica de la Materia Condensada, C-III.
Universidad Aut\'onoma de Madrid, 28049 Madrid, Spain.} 

\maketitle

\begin{abstract}
\noindent
We present a systematic Density Functional
Theory (DFT) study of geometries and energies of the nucleic
acid DNA bases (guanine, adenine,  cytosine and thymine) 
and 30 different DNA base-pairs. 
We use a recently developed linear-scaling DFT scheme,
which is specially suited for systems with large numbers
of atoms. 
As a first step towards the study of large DNA systems, in this
work: (i) We establish the reliability of the approximations
of our method (including pseudopotentials and
basis sets) for the description of the hydrogen-bonded
base pairs, by comparing our results with 
those of former calculations. We show that 
the interaction energies at Hartree-Fock geometries
are in very good agreement with those of 
second order M{\o}ller-Plesset (MP2) perturbation 
theory (the most accurate technique that can be
applied at present for system of the sizes 
of the base-pairs).
(ii)  We perform DFT structural optimizations 
for the 30 different DNA base-pairs,
only three of which had been previously studied
with DFT.
Our results provide information on the effect of correlation
on the structure of the other 
27 base pairs, for which only Hartree-Fock
geometries were formerly available.
\end{abstract}

]

\narrowtext

\section{Introduction}

Hydrogen bonds between complementary
purine-pyrimidine bases play a significant
role in the bonding between the two 
strands of the double helix structures of 
DNA~\cite{Watson}.
Nevertheless, other factors are also of paramount importance
in determining the structure of the helix. For instance,
base-stacking~\cite{Sponer4,Sponer5,Sponer6} 
plays a crucial role
in preserving the hydrophobe aromatic rings from 
interacting with water molecules, besides contributing
to increase the Van der Waals interactions.
Also, both counterions~\cite{Burda1,Burda2} and 
water molecules are important
in screening the electrostatic repulsion
between the negatively charged phosphate groups. 
The theoretical study of these systems and the
effects of each type of interaction have been 
hindered by the great complexity of the calculations,
both due to the difficulty in treating such 
different interactions, and to the large 
number of atoms involved. 

Although some progress has been done recently,~\cite{Brameld}
semi-empirical quantum chemistry (QC) methods
and empirical force fields~\cite{Hubalek,Kabelac} 
are generally not accurate enough to describe these systems.
The most reliable procedure is undoubtedly the {\em ab-initio}
QC approach, in which the accuracy of the calculation can 
be systematically improved by increasing the size and quality of
the basis set and the level of correlation included~\cite{Sponer3}.
This can be a powerful method to study DNA and other
biological systems, and a unique tool to address some of their properties.  
However, the intensive numerical effort required by these
methods poses a serious limitation to the system sizes
that can be handled at a satisfactory level of the
theory~\cite{Sponer3}, a fact which has precluded
their use in realistic biological systems. 

An alternative, which allows calculations
for system sizes well beyond the limits 
of the traditional 
{\em ab-initio} QC methods, is the Density Functional 
Theory~\cite{Hohenberg,Kohn} (DFT).
At present, it appears that DFT 
is the only first principles method
potentially capable of handling the large sizes involved, 
although the standard DFT techniques are still too expensive 
to solve systems with more than 
a very few hundreds of atoms, like
those in DNA molecules. If the DFT methods are
to make an impact in biological systems, it is
neccesary to be able to go beyond the current
size limits, but maintaining the current accuracy.
In this context, an extremely promising
development has been the recent search
for computational techniques in 
which the numerical effort scales only
linearly with system size: the so called 'order-N' 
methods (see Ref.~\onlinecite{Ordejon} for a review). 
They open, for the first time, the possibility
of performing calculations in very large molecules,
and have already been applied 
to the study of DNA chains with many hundreds
of atoms by means of semi-empirical Hamiltonians~\cite{York}
and approximate, non-selfconsistent DFT~\cite{Lewis}.

The application of order-N techniques in the context
of fully first-principles, selfconsistent DFT calculations
is, in general, in a less advanced development state. 
Nevertheless, we have recently proposed
a DFT method and the corresponding computer code SIESTA,
with order-N scaling, which is able to do
such calculations in systems with thousands of atoms,
in single-processor workstations~\cite{ordejon1,ordejon2,ijqc,Daniel}.
Our preliminary tests~\cite{ordejon2,ijqc} 
have demonstrated that the method
is able to treat systems as large as a whole turn of a DNA
chain with more than 650 atoms, therefore
opening the possibility of 
studying large biological molecules
from first principles.  Work in this direction is 
underway, and will be published elsewhere.\cite{Artacho}

The purpose of this paper,
as a first step in the application of the method to 
complex biological systems, and in particular to DNA molecules,
is to make a thorough study of isolated nitrogenated bases and
hydrogen-bonded base-pairs. This study serves
to validate the present method (basis orbitals, approximations and
numerical techniques) for the study of these base-pairs,
by comparing the energies and geometries with those
of previous calculations, where available.
The results presented here show that our method provides
a very accurate description of these systems, with the advantage of being
considerably fast and, as mentioned, capable of reaching 
very large system sizes.  
Besides, we provide a systematic DFT study of the structures of
the different base-pairs, and the effect of the relaxations
on the interaction energies. The use of our DFT scheme
to obtain equilibrium geometries has the advantage that
it includes correlation effects, which are absent
in the available Hartree-Fock (HF) geometries.\cite{Sponer96}  
At the same time, it is computationally feasible, 
unlike the M{\o}ller-Plesset second order 
perturbation theory~\cite{MP2} (MP2) method.

The rest of this paper is organized as follows.
In Section \ref{sec:method} we discuss the details
of our DFT method and of the calculations performed.
Section \ref{sec:HF} describes the 
energetics of the base-pairs at the available HF geometries,
comparing our results with those of MP2 calculations in the
literature.
Section \ref{sec:DFT} presents our results
for the DFT structural relaxations. Finally, in
Section \ref{sec:conclusions} we present the conclusions
of this work.

\section{Method and calculations details}
\label{sec:method}

\subsection{The SIESTA method}
All our calculations have been done 
with SIESTA~\cite{ijqc}, a code for DFT calculations in 
systems with a large number of atoms, in which
the cost of the calculation (both in memory and
CPU time) scales linearly with the size 
of the system. Here we 
give only a brief description of the basic
approximations involved in the calculation, whereas
a detailed description can be found in Refs.~\onlinecite{ijqc,Daniel}.

We treat exchange-correlation (XC) within the framework of 
the Kohn-Sham formulation~\cite{Kohn} of DFT~\cite{Hohenberg}.
It is rather well known from many calculations in a 
variety of systems~\cite{sim,laasonen,barnett,hamann,Santamaria} that
a correct description of hydrogen-bonds requires the
use of non-local XC functionals. The Local Density Approximation
(LDA) yields bond distances in the hydrogen bonds
which are about 10$-$15\% shorter and binding energies 
about 50$-$70\% larger than the experimental values.
Inclusion of gradient corrections in several
Generalized Gradient Approximations (GGA) functionals
improves the description dramatically, achieving levels of 
accuracy an order of magnitude better than LDA.
In this work we have 
used the first principles GGA functional proposed 
recently by Perdew, Burke and Ernzerhof~\cite{Perdew} (PBE).

SIESTA  uses non-local, norm-conserving pseudopotentials to 
eliminate the core electrons from the calculation,
and to produce a smoother valence charge density.
In this work, the pseudopotentials are
obtained from first principles, following
the scheme of Troullier and Martins~\cite{Troullier}.
The valence electrons are described using 
the linear combination of atomic orbitals (LCAO)
approximation.

An essential ingredient for the
linear scaling within this approach is the finite range
of the matrix elements between atomic orbitals. 
To achieve it, we use basis orbitals which strictly
vanish beyond a cutoff radius~\cite{eshift} 
(instead of the usual approach 
of using decaying orbitals and neglecting matrix elements
by  whatever criterion).
The main advantage is consistency: given a basis,
the eigenvalue problem is solved for the {\em full} Hamiltonian.
Thus, the procedure is  numerically very stable even for
short ranges, in contrast with the usual approach.
Since the computational load grows substantially
with the basis range,
it is important to work with basis functions 
that display fast convergence for short orbital ranges.
We have developed a scheme for finite range basis
set generation which we will now outline.~\cite{Daniel}

In this and previous works, the radial parts of the
finite-range orbitals were determined in the spirit 
of the method of Sankey and Niklewski,~\cite{Sankey} who proposed a 
scheme for minimal (single-$\zeta$) bases that
we have generalized to arbitrarily complete sets.
The single-$\zeta$ orbitals are obtained
by solving the DFT atomic problem
(including the pseudopotential) with 
the boundary condition for the orbitals
of being zero beyond the cutoff radius, 
while remaining continuous.
For the efficient generation of larger, more 
complete basis sets we have used the ideas 
developed within the QC community over the years,
incorporating them into new schemes adapted 
to numerical, finite-range bases for linear scaling.
Numerical multiple-$\zeta$ bases are 
constructed in the split-valence
philosophy~\cite{ijqc,Daniel}. Given an atomic orbital, 
it is split into two or more functions.
The first splitting is made by 
introducing a smooth function that
reproduces exactly the tail of the 
original orbital beyond a specified
radius. The difference between the 
original orbital and this smooth
function is an orbital with an even 
shorter range. Multiple splits are
obtained by repeating the procedure.
Our approach also allows polarization 
orbitals~\cite{Daniel}. These are obtained by
numerically solving the problem 
of the isolated atom in the presence of a
polarizing electric field.
Comparing the solution with a perturbative
expansion (Sternheimer equations~\cite{Mahan})
gives the shape of the wanted polarization orbitals.
The cutoff radius of the polarization orbitals
is therefore the same as the one of the shell 
being polarized.

In all the calculations presented in this work we have used a 
double-$\zeta$ (split valence) basis with polarization functions 
in all the atoms (including hydrogen).
The cutoff radii for the atomic orbitals of 
each element can be seen in Table~\ref{radii}, as were 
obtained by fixing a confinement energy~\cite{eshift} of 50 meV.

The matrix elements of the different terms of the Kohn-Sham
Hamiltonian are calculated in one of two different 
ways~\cite{ijqc}.
The terms that involve integrals over two atoms only (kinetic energy,
overlap, and other terms related 
with the pseudopotential, see below) are 
computed a priori as a function of the
distance between the centers, and stored in tables 
to be interpolated later with very little use 
of time and memory. The other terms 
are calculated with the help of a 
uniform grid of points in real space.
The smoothness of the integrands
determines how fine a grid is needed, 
and, of course, the finer the grid, the 
more expensive the calculation. 
We remark that the use of pseudopotentials, which eliminates
the rapidly varying core charge, is essential 
to provide functions smooth enough to make the grid
integration feasible. This fineness is measured by the energy
of the shortest wave-length plane-wave that can be described with the
grid, in analogy with plane-wave calculations.
In all the calculations presented here, we have used a 
cutoff of 125 Ry.
   
The calculation of the pseudopotential matrix elements
is done very efficiently using the
Kleinman-Bylander~\cite{Kleinman} factorized form. 
It allows the three-center integrals of the pseudopotential between
atomic orbitals to be treated as products of 
two-center integrals, which are tabulated as described
above.

With the bases, approximations and techniques described
so far, the Kohn-Sham hamiltonian is built up with 
order-N operations. 
The solution to the eigenvalue problem can also be obtained
with a linear scaling effort using techniques recently
developed, and available in SIESTA~\cite{Ordejon,ordejon3}. 
For the small systems considered here, however, the
straight diagonalization (which scales as the cube of the number
of orbitals) requires very little effort, 
and therefore has been used
in this work to solve the Kohn-Sham eigenvalue equations.

\subsection{Details of the calculations}

In order to reach reliable conclusions 
about the accuracy of our method, 
we have used in this study a large set of 30 base-pairs.
Besides the common Watson-Crick guanine-cytosine 
and thymine-adenine pairs, we also
consider a significant range of other configurations
of the four bases guanine, cytosine, 
adenine and thymine (G, C, A, T).
These are the same as those studied 
by $\rm{\check{S}poner}$ {\em et al.}~\cite{Sponer96}
in their MP2 study. The Watson-Crick configurations are
designated WC, and the Hoogsteen, 
reversed Hoogsteen and reversed 
Watson-Crick appear as H, RH, RWC 
respectively. Other configurations 
are distinguished simply with numbers, 
eg. AA1, AA2, etc.   
In assigning the numbers to the pairs we
have followed the nomenclature 
of Hobza and Sandorfy~\cite{Hobza}, who classified
the pairs in order of decreasing stability.
Their ordering was not confirmed by
later results (including ours), but the
convention is nevertheless maintained to simplify
comparisons and avoid confusion. 
The structures of the bases and base-pairs
studied in this work can be found in Figures
1 and 2 of Ref.~\onlinecite{Sponer96}.
For the numbering of the atoms we followed
Ref.~\onlinecite{Sponer2}.

In the calculation of the energetics of the base-pairs,
we have analyzed two different quantities. 
First, the interaction
energy $E_{int}$, defined as the energy of the base-pair
minus the energy of each base with the same
geometry it has in the pair.
Second, the total stabilization energy,
$E_t$, defined as the difference between the energy of the pair 
and that of each base in its isolated optimal geometry. 
Therefore, the difference between $E_t$ and $E_{int}$
is the deformation energy, {\em i.e.}, the increase in 
intramolecular energy due to the geometry change
when the base-pair is formed.

Due to the finite size of the bases used, 
both $E_{int}$ and $E_t$ have to be 
corrected for the basis set superposition
error~\cite{Boys} (BSSE).
In this work, all the energies have been corrected 
for BSSE, as described in the following.
For the interaction energy, we have used the standard 
Boys-Bernardi counterpoise correction~\cite{Boys}:
the BSSE is calculated as the difference
between the energies of the isolated bases
obtained with the orbitals of the base alone,
and with the ``ghost" orbitals of the other base: 
$ BSSE = E(A) + E(B) - E(A*) - E(B*) $
(where the asterisks indicate the inclusion
of the orbitals of
the other base in the calculation).
The same correction is used for the
stabilization energy $E_t$. Since
$E_t$ contains the deformation energy $E_{def}$,
this approach is valid only if $E_{def}$ is not
much affected by the BSSE ({\em i.e.}, if the
change in the BSSE is not large 
when calculated with the relaxed 
isolated bases geometry instead
of the coordinates of each base in the 
pair)~\cite{Sordo}. We tested 
this and found that the variation in BSSE calculated 
with these two geometries is only 
about 10\% of the total
BSSE value, so we will consider that the BSSE correction
defined above is as valid for $E_t$ as for $E_{int}$.    

The structural relaxations were done by means
of a conjugate gradient minimization
of the energy, until the forces on all atoms
were smaller than 0.04 eV/\AA.  No constraints were imposed in
the relaxation, except the planarity of the base-pairs. 
This constraint was imposed in order to facilitate comparison
with the results of $\rm{\check{S}poner}$ {\em et al.}, who also 
analyze planar bases and base-pairs.
In the relaxations, forces are calculated as analytical
derivatives of the total energy~\cite{ijqc}. 
No BSSE correction was included in the forces.
This would lead to problems if the BSSE had an important
variation with atomic positions, since in that
case the relaxed geometries obtained without the BSSE
correction would not correspond to the minimum of
the total energy including the BSSE correction.
In order to 
check this, corrected and not corrected
interaction energies of the AA1 base-pair 
were calculated as a function of the distance,
separating the molecules rigidly in the 
H bonds direction. 
As expected, the BSSE is more important in absolute 
value as the two bases are brought closer,
and decreases as the bases are separated. 
However, the results show that
this variation does not appreciably affect neither 
the equilibrium distance (the acceptor-hydrogen
distance was 1.95 \AA~for the 
uncorrected curve and 1.97 \AA~for the corrected one)
nor the vibrational frequencies.  

Finally, it is worth mentioning the computational 
requirements of the calculations presented here.
All the computations were performed in a PC with
a 120 MHz Pentium processor. For the relaxations
of the base-pairs,
the calculations took an average of about 4 hours
of CPU time for each relaxation step. The memory
usage was below 100 Mb, and virtually no disk
use was neccesary (all the integrals being stored
in memory). This shows
the efficiency of the code, and the possibility 
of studying relatively large systems in very
modest platforms.

\section{Energies at HF/6-31G** geometries}
\label{sec:HF}

In order to demonstrate the accuracy and validity of
our method for the description of the energies of hydrogen-bonded
base-pairs, we need a reference with which to
compare our results. Since the experimental information
on energies and structures of isolated 
bases and base-pairs is
very scarce, we have used the results from 
former {\em ab-initio} calculations as a benchmark. 
There is a large amount of work done in these
systems in the context of {\em ab-initio} QC methods.~\cite{Sponer3}
Probably the most complete and sophisticated calculations
are those performed by $\rm{\check{S}poner}$ and coworkers,\cite{Sponer96} 
using the MP2 method.
This usually covers a substantial part of the
correlation energy, and is the most accurate correlation 
technique that can be applied at present for systems of
the size of a few tens of atoms, like the base-pairs.
However, these calculations are still computationally expensive,
and therefore they can only be done using medium sized bases
(typically 6-31G**) and fixed geometries
obtained with simpler schemes like HF.
Geometry optimizations at the MP2 level 
have only been possible for the smallest, highest
symmetry base-pair (cytosine-cytosine)\cite{Sponer96}.

We will therefore discuss the interaction energies obtained
with SIESTA for the base-pairs in the HF/6-31G** geometries of
$\rm{\check{S}poner}$ {\em et al.}~\cite{Sponer96}, and
compare with the corresponding MP2 results.
The data are shown in Table \ref{eint-hf}. 
GG2 and GC2 base-pairs are not included in this
table because at this level of relaxations 
these pairs are not stable and 
converge to the configurations of GG1 and GCNEW, respectively.
We compare with the MP2 results~\cite{Sponer96} 
evaluated on the same geometry. We also show the 
percentage deviation between both results.
For all the base-pairs except GG4, the
agreement is considerably good, with differences 
smaller than 8\% and much less in most cases.
GG4 seems to be an exception to the general 
trend as its difference with the 
MP2 value is 26\%. We tried to see if this 
was a problem of the basis set and 
made calculations with larger cut-off 
radii for the atomic orbitals. 
For an energy shift of 10 meV the interaction
energy was -8.4 kcal/mol, so the deviation with 
respect to the MP2 results is reduced to 16\%,
but it is still far larger than for the rest
of the base-pairs (for other
base-pairs, the difference in the interaction energy
for the 50 and 10 meV energy shift bases is much smaller
than in the GG4 case).
  
The standard deviation of our results compared to the MP2
values is of 0.73 kcal/mol. It is interesting to compare
these results with the DFT values obtained by $\rm{\check{S}poner}$
{\em et al.}\cite{Sponer96} for the same HF/6-31G** 
geometries, using the Becke3LYP functional~\cite{b3lyp}. 
The largest deviation of their results is
of 1.3 kcal/mol (11 \% of $E_{int}$) for the TC1 pair,
while the standard deviation from the MP2 results
is accidentally the same as ours: 0.73 kcal/mol.
We can therefore conclude that the results obtained
for the energies at fixed geometries 
using the PBE functional 
are of similar degree of accuracy as those obtained
by other authors with other GGA functionals.
The standard deviation between our results
and the DFT data of $\rm{\check{S}poner}$ {\em et al.} is about
0.85 kcal/mol, of the same order 
as the difference with the
MP2 results. This  serves to validate the PBE functional,
as well as SIESTA and the approximations involved 
(cut-off bases, pseudopotentials, grid integrations, etc),
as a valuable and competitive tool compared with standard,
all-electron, gaussian-bases DFT programs.

The current LDA and GGA implementations of DFT 
are not able to describe accurately Van der Waals or
dispersion interactions.
Still, the previous results show that the common non-local
XC functionals provide quantitatively accurate values
of the interaction energies of the 
hydrogen-bonded base-pairs.
Although the reason for this is not fully clear
yet, it seems that the dispersion energies in 
these H-bonded systems are significantly smaller than
the ones that would be predicted using an 
empirical London dispersion energy~\cite{Sponer96}.
It is still unclear if current XC functionals
are able to describe the interaction between
stacked base-pairs, where the dispersion energy is larger
than expected from an empirical London formula.
Calculations by $\rm{\check{S}poner}$ and coworkers~\cite{Sponer6}
seem to indicate that
the Becke3LYP functional significantly underestimates
the dispersion energies for stacked base pairs. A study of the performance 
of other XC functionals with our method is underway, 
and will help in clarifying this issue.

To conclude this section, we show in Table~\ref{dipole-hf}
the dipole moments for the HF/6-31G**
geometries. We compare the results of the HF/6-31G** calculations
of $\rm{\check{S}poner}$ {\em et al.}~\cite{Sponer96}
with those of this work. 
DFT provides lower values, due to the tendency
of the Hartree-Fock approximation
to overestimate the electrostatic interactions.

\section{DFT geometry optimizations} 
\label{sec:DFT}

At present, DFT studies of the geometries of
base-pairs are rather scarce. To our
knowledge, only the geometries of the
Watson-Crick TA and GC, and the CC
base-pairs have been obtained with
DFT. Therefore, we have 
extended the previous DFT works, and performed a
thorough study of  the energetics and structural
relaxations of nitrogenated base-pairs with our
DFT approach.
Here we describe the results 
of such structural optimizations for the 30 hydrogen-bonded
base-pairs (27 of which had not been studied with DFT previously).

\subsection{Isolated bases}

We have first optimized the geometries of isolated 
adenine, guanine, cytosine and thymine using our method. 
The results for the geometries obtained (bond distances
and angles) are presented in Tables~\ref{ade}-\ref{thy}.
For comparison, we also show the 
DFT results of Santamar{\'{\i}}a
and V\'azquez~\cite{Santamaria} (S-V), 
obtained with the Vosko-Wilk-Nusair~\cite{vwn} 
functional with Becke-Perdew~\cite{bp} 
non-local corrections, as well as the
experimental values obtained with X-ray diffraction 
for crystallized DNA~\cite{Taylor,Ozeki}.
The bases were relaxed for planar geometries. 

We can observe that the geometries obtained here
are close to those of S-V with an all-electron calculation
with gaussian bases and a different XC functional.
Again, this supports the reliability of SIESTA and
its approximations. The bond distances obtained in this work
are usually slightly larger than the values of S-V, 
although the largest difference is only 0.016 \AA.
The results of S-V are slightly closer to the  
experimental data. These must be taken only
as a rough reference, since they correspond to
measurements of DNA crystals; packing forces 
constrain the molecules, so all the bond distances 
are shorter than those calculated for the free bases.
Some of the differences between our results and those
of S-V are due to the restriction of planarity in
our calculation, which was not imposed by S-V.
It is well known that the amino groups of 
the nucleic acid bases suffer a
pyramidalization due to partial $sp^3$ 
hybridization~\cite{Sponer3,Sponer2}: 
the two H atoms go out of the plane of the aromatic ring
whereas the N atom moves in the opposite direction. 
Therefore, there are in some cases important differences 
in the distances and angles that involve 
atoms in the amino groups of the bases between
the planar and non-planar bases.

\subsection{Base-pairs}

Here we discuss our results for the structural optimizations
of the base-pairs performed with the SIESTA program. 
Our relaxations start from HF/6-31G** 
geometries, with the
exception of GG2 and GC2 base-pairs, 
(not stable at the HF/6-31G** level)
for which we start from the HF/MINI-1 coordinates.
In all cases, planar symmetry was imposed in the
relaxations.

Table~\ref{bases-relax} shows the H-bonds distances 
and angles for the base-pairs obtained with our method.
Donor-acceptor and donor-hydrogen distances are shown, 
together with the angle subtended by the
three atoms involved in the bond.          
Comparing with HF/6-31G** geometries (see Ref.~\onlinecite{Sponer96}), 
we see that hydrogen bridges are shorter and 
donor-H distances are larger in our calculations.  

Among the hydrogen bonds in the base-pairs studied, 
33 are N(H)$\cdot\cdot\cdot$N and 29 are O$\cdot\cdot\cdot$(H)N. Their distances range from
2.755 to 3.169 \AA. Among the longest of them,
there is a clear majority of N(H)$\cdot\cdot\cdot$N, which seems logical 
because the electrostatic attraction
between the H atom and a N atom should 
be weaker than the attraction 
between H and O. However, there are also
some N(H)$\cdot\cdot\cdot$O bonds that are quite long.
The reason is that several factors and interactions, 
and not only the atoms which
are involved in the H-bond, influence the final configuration 
of the pairs.  

Distances between donor and H atoms range from 1.026 to 1.070 \AA. 
There is in many cases a correlation between short H-bond distances
and long donor-H distances. It is clear that the greater the electrostatic
attraction the H atom suffers from the other base, the longer will be
its distance from the donor atom and vice-versa. However, again
this cannot be taken as a strict rule, as the final position of 
each atom is determined by all the neighboring atoms.

The dipole moments at the geometries relaxed with
our approach
are shown in Table~\ref{dipole-hf}.
They are all smaller than the HF/6-31G** values, except
for two of the bases, and 
there are no major differences with the results obtained
with SIESTA for the HF/6-31G** geometries, 
although there is in almost all cases a slight increase
in their values.

Table~\ref{ener-relax} shows the interaction ($E_{int}$) and total 
stabilization ($E_T$) energies for the base-pairs, obtained
at the SIESTA relaxed coordinates.
The ordering of the base-pairs is the same as the
one of Table~\ref{eint-hf}, to facilitate comparison.
Interaction energies range from -32.2 to -9.6 kcal/mol, and 
stabilization energies from -27.7 to -8.0 kcal/mol. 
The most stable base-pair is GCWC, in agreement with previous 
results~\cite{Sponer96,Hobza}, and the ordering of the next three 
base-pairs is also the same.
The relative ordering (according to $E_{int}$) of all the TA, 
GT, GG, GA, AA and TT tautomeres is conserved, 
but AC2 is more stable than AC1 and TC2 more stable than TC1
with our method. 
It is interesting to see that the energetic ordering 
depends on which of the two energies is used, $E_{int}$ or $E_t$. 
This is not the case in MP2//HF results~\cite{Sponer96}, where 
only two base-pairs change position when ordered according to 
$E_{int}$ instead of $E_t$ (although the ordering could
change further if the geometries were obtained at the MP2
level, too).

Several points are worth noticing from the
results of Table~\ref{ener-relax}. 
(i) The interaction energies are systematically
larger (by a few kcal/mol) than those obtained, 
with the same theory, at the HF/6-31G** 
geometries (compare with the results of Table~\ref{eint-hf}).
This indicates that the geometry of the 
H-bonded base-pairs configurations are
more sensitive to the details of the calculation than those
of the free bases, and that the HF/6-31G** geometries for
the H-bonds are not fully optimal for DFT calculations. 
(ii) The deformation energies (difference 
between $E_T$ and $E_{int}$) are 
larger than those obtained at the HF/6-31G** and MP2 levels
for HF/6-31G** geometries. In our calculations, the deformation
energy ranges up to as much as 6.7 kcal/mol for the GCNEW pair.
The same trend is observed in other DFT calculations
(see below). (iii) The GG2 and GC2 base-pairs, which at the 
HF/6-31G** are unstable and converge towards the GG1 and GCNEW 
respectively, are found to be stable in our geometry
optimization, which was started from HF/MINI-1 coordinates. 
The interaction energy of these pairs is small, but comparable
to many of the other base-pairs.

To our knowledge, DFT optimizations of 
base-pairs by other authors are only
available for GCWC,TAWC and CC~\cite{Sponer96,Santamaria}. 
Tables~\ref{gcwc},~\ref{tawc} and \ref{cc} present
the available DFT results together with those
of SIESTA and HF/6-31G**, for these three pairs. 
For CC, results of a MP2 geometry optimization are also
available from the work of $\rm{\check{S}poner}$ {\em et al.}~\cite{Sponer96},
and are included in Table~\ref{cc}.
Acceptor--donor and donor--hydrogen distances are shown, 
together with the interaction and stabilization energies.
In general, our results report donor-acceptor 
distances which are only slightly shorter than those
of other DFT calculations (with a maximum difference 
of 0.058 \AA), whereas the donor-H distances
are in excellent agreement. All DFT results yield
shorter D-A distances than the HF/6-31G** approximation.
For the energies, the dispersion in the DFT reports
is considerable. Our results agree quite well
with those of $\rm{\check{S}poner}$ {\em et al.}~\cite{Sponer96}
(available for GCWC and CC), whereas the differences
with those of Santamar{\'{\i}}a and V\'azquez~\cite{Santamaria}
(available for GCWC and TAWC) are larger.
We note that, as mentioned before, deformation energies
are considerably larger in all the DFT results than in the 
MP2//HF results. For instance, for GCWC, the DFT calculations
yield values of the deformation energies of 3.3, 4.8 and 4.7 kcal/mol,
whereas the MP2//HF result is of only 2.1 kcal/mol.

It is interesting to discuss the case of CC, since it is the only
base-pair where coordinates optimizations at the MP2 level
are available. We see in Table~\ref{cc} that the bond length
between the donor and acceptor atoms is underestimated by the
DFT calculations, and overestimated by the HF/6-31G**
by about the same amount. However, it seems that
the DFT energies are closer to the MP2 values at the
MP2 geometries than at the HF ones. 
Also, the MP2 deformation energy increases 
from 1.3 kcal/mol for the HF/6-31G** geometries
to 1.8 kcal/mol for the MP2 geometries, and therefore
approaching that obtained with DFT (2.3 kcal/mol in the
results of $\rm{\check{S}poner}$ {\em et al.} 
and 2.4 kcal/mol in our case).

\section{Conclusions}
\label{sec:conclusions}

In this work, we have performed DFT calculations
on the DNA bases adenine, guanine, cytosine and thymine,
and 30 base-pairs formed by these bases. 
The calculations were performed with
the SIESTA code, which is a novel technique
for DFT calculations in systems with large
numbers of atoms using pseudopotentials
to describe the effect of the core electrons and
finite range basis orbitals for the valence electrons.
The calculations presented here serve to validate
our method for the study of H-bonded base-pairs,
as a first step toward the complete DNA helix
(which is feasible with SIESTA due to the linear
scaling of the numerical effort with the number
of atoms in the system).

For calculations on the HF/6-31G** geometries,
excellent agreement with MP2 results 
was obtained. The deviations are smaller
than 8\% (which amounts to 1.3 kcal/mol at most),
except for GG4,which  differs quite 
significantly from MP2 results. Calculations
with longer atomic orbital radii reduce
the difference, but it is still bigger than the
rest. The dipole moments for these geometries
are systematically lower than those of the
HF calculations.

For the isolated bases, the planar geometries
obtained in our calculation are in good 
agreement with former DFT results.

The relaxed geometries of the 30 DNA base-pairs
were also obtained with our method.  The donor-acceptor
distances in the hydrogen bonds are systematically
shorter than those of HF/6-31G**, as in other
DFT calculations~\cite{Santamaria,Sponer96}.
Our results compare
well with other DFT optimizations of the GCWC and
TAWC base-pairs.
For the CC pair, for which MP2 optimizations are available,
the results of SIESTA and other DFT calculations
slightly underestimate the hydrogen bond distances,
but provide a quite accurate value for the interaction
energies. The deformation energies upon the dimer formation
are larger for the DFT results than for the HF geometries,
a result that is in agreement with the increase of $E_{def}$
in the MP2 approximation when MP2 coordinates are used
in the calculation.

Dipole moments for the relaxed geometries 
are quite similar to  
our previous results for HF geometries, but
slightly larger in most cases.

The results for the energetic ordering of the base-pairs
have also been analyzed. Although there are not
essential changes, the ordering is slightly
different than for the MP2//HF results.
However, the relative order between tautomers
is conserved in most cases.
The GG2 and GC2 base-pairs, which were unstable
at the HF/6-31G**  level, are stable in our
calculations, and have interaction energies similar
to the other base-pairs.

In conclusion, the results presented here show
that SIESTA is a valuable tool for the study
of H-bonded DNA base-pairs. It provides results
very similar to other DFT techniques, and which
compare very well with the available MP2 data.
Work is under progress to determine the
validity of the method for the properties
of stacked bases, and for the study of large DNA segments.

{\bf Acknowledgments}.
We are grateful to J. $\rm{\check{S}poner}$ and R. Santamar{\'{\i}}a 
for making their coordinates of
bases and base-pairs available to us.
We are also indebted to R. Weht for many useful discussions and
his help during the first stages of this work.
This work was supported by Spain's DGES under grant PB95-0202.
P. O. was partially supported by a Sponsored Research
Project from Motorola Phoenix Corporate Research Laboratories.


\hfill\break
\clearpage

\begin{table}
\caption{Atomic Orbitals radii (Bohr) for an energy shift of 50 meV.
For each L shell, $\zeta$ stands for each of the split valence orbitals.}
\label{radii}
\normalsize
\begin{tabular}{ccccc}
        & \multicolumn{2}{c}{ L= 0 }  &  \multicolumn{2}{c}{ L= 1} \\  
Species  & \footnotesize $\zeta$=1 & \footnotesize $\zeta$=2 & 
\footnotesize $\zeta$=1 & \footnotesize $\zeta$=2   \\ 
\tableline
         H 
&          6.047 
&        2.488 
&        
& \\

         C 
&        4.994 
&        3.475 
&          6.254
&          3.746\\

       N 
&      4.390 
&        2.942 
&          5.496 
&          3.092 \\

         O 
&        3.937 
&        2.542 
&         4.931  
&          2.672 \\
\end{tabular}
\normalsize
\end{table}

\begin{table}
\caption{Base-pair Interaction Energies ($E_{int}$, in  kcal/mol)
at HF/6-31G** geometries.}
\label{eint-hf}
\begin{tabular}{cddd}
Pair & MP2$^{\rm a}$    & SIESTA  & Deviation (\%) \\
\tableline
GCWC  & -25.8 & -26.8 & -3.9 \\
GG1   & -24.7 & -25.1 & -1.6 \\
GCNEW & -22.2 & -21.7 &  2.2 \\
CC    & -18.8 & -17.5 &  6.9 \\
GG3   & -17.8 & -16.6 &  6.7 \\
GA1   & -15.2 & -15.5 & -2.0 \\
GT1   & -15.1 & -15.0 &  0.7 \\
GT2   & -14.7 & -14.5 &  1.4 \\
AC1   & -14.3 & -14.0 &  2.1 \\
GC1   & -14.3 & -14.7 & -2.8 \\
AC2   & -14.1 & -14.7 & -4.2 \\
GA3   & -13.8 & -13.8 &  0.0 \\
TAH   & -13.3 & -13.7 & -3.0 \\
TARH  & -13.2 & -13.6 & -3.0 \\
TAWC  & -12.4 & -12.3 &  0.8 \\
TARWC & -12.4 & -12.3 &  0.8 \\
AA1   & -11.5 & -11.7 & -1.7 \\
GA4   & -11.4 & -11.7 & -2.6 \\
TC2   & -11.6 & -10.8 &  7.5 \\
TC1   & -11.4 & -10.6 &  7.0 \\
AA2   & -11.0 & -11.4 & -3.6 \\
TT2   & -10.6 & -9.9  &  6.6 \\
TT1   & -10.6 & -10.1 &  4.7 \\
TT3   & -10.6 & -10.2 &  3.8 \\
GA2   & -10.3 & -10.6 & -2.9 \\
GG4   & -10.0 & -7.4  & 26.0 \\
AA3   & -9.8  & -9.8  &  0.0 \\
2aminoAT & -15.1 & -15.2 & -0.7
\end{tabular}
\noindent
$^{\rm a}$ From ref.~\onlinecite{Sponer96}
\end{table}


\begin{table}
\caption{Dipole moments (Debyes) of the DNA base-pairs$^{\rm a}$}
\label{dipole-hf}
\begin{tabular}{ldddd}
Pair &  HF//HF &  SIESTA//HF & 
Difference (\%) & SIESTA \\
\tableline
GCWC & 6.5  & 5.8  & -10.8 & 6.1 \\
GG1  & 0.0  & 0.0  & 0.0  & 0.0 \\
GCNEW& 3.1  & 3.4  & 9.7 & 3.3 \\
CC   & 0.0  & 0.0  & 0.0  & 0.0 \\
GG3  & 10.5 & 10.3 & -1.9  & 10.9 \\
GA1  & 5.6  & 4.7  & -16.1 & 4.9 \\
GT1  & 7.7  & 6.9  & -10.4 & 7.0 \\
GT2  & 8.6  & 8.0  & -7.4  & 8.3 \\
AC1  & 4.8  & 3.5  & -14.6 &  3.7 \\
GC1  & 12.7 & 10.7 & -15.7 & 11.5 \\
AC2  & 9.7  & 8.3  & -14.4 & 8.6 \\
GA3  & 8.8  & 7.9  & -10.2 & 8.4 \\
TAH  & 6.4  & 5.5  & -14.1 & 5.7 \\
TARH & 5.9  & 5.0  & -15.2 & 5.0 \\
TAWC & 2.0  & 1.4  & -30.0 & 1.4 \\
TARWC& 2.5  & 2.3  & -8.0  & 2.4 \\
AA1  & 0.0  & 0.0  & 0.0  & 0.0 \\
GA4  & 9.2  & 8.2  & -10.9 & 8.8 \\
TC2  & 4.5  & 3.9  & -13.3 & 3.8 \\
TC1  & 5.9  & 5.3  & -10.2 & 5.5 \\
AA2  & 4.9  & 4.7  & -4.1  & 4.8 \\
TT2  & 0.0  & 0.0  & 0.0  & 0.0 \\
TT1  & 1.3  & 1.3  & 0.0  & 1.6 \\
TT3  & 0.0  & 0.0  & 0.0  & 0.0 \\
GA2  & 7.3  & 6.4  & -12.3 & 6.8 \\
GG4  & 0.0  & 0.0  & 0.0  & 0.0 \\
AA3  & 0.0  & 0.0  & 0.0  & 0.0 \\
2aminoAT & 4.2 & 4.0 & -4.7 & 4.2 \\
GG2  & --   & --   & --   & 12.7 \\
GC2  & --   & --   & --   & 13.8\\
\end{tabular}
\noindent
$^{\rm a}$ HF//HF: Hartree-Fock results obtained at
HF/6-31G** coordinates. From ref.~\onlinecite{Sponer96}.

\noindent
SIESTA//HF: results of this work, obtained at 
HF/6-31G** geometries.

\noindent
Difference: Percent difference between HF//HF and SIESTA//HF
results.

\noindent
SIESTA: results of this work, calculated at the
SIESTA relaxed coordinates.
\end{table}

\hfill\break
\clearpage

\widetext

\begin{table}
\caption{Bond distances and angles for isolated adenine$^{\rm a}$.}
\label{ade} 
\begin{tabular}{cccccccc}
Distances (\AA) & This work & Ref.~\onlinecite{Santamaria} 
& Exp. & Angles (deg)
& This Work & Ref.~\onlinecite{Santamaria} & Exp.\\
\tableline
{C8-N9} & 1.391 & 1.387 & 1.367  & {C8-N9-C4} & 107.30 & 106.74 & 105.9\\
{N9-C4} & 1.390 & 1.386 & 1.376 & {N9-C4-C5} & 103.67 & 104.50 & 105.7\\
{C5-N7} & 1.394 & 1.394 & 1.385 & {C5-N7-C8} & 103.35 & 103.75 & 103.9 \\
{N7-C8} & 1.334 & 1.324 & 1.312 &  {N7-C8-N9} & 113.40 & 113.49 & 113.8\\
{C4-N3} & 1.355 & 1.349 & 1.342 & {C5-C4-N3} & 127.53 & 126.98 & 126.9\\
{N3-C2} & 1.357 & 1.348 & 1.332 & {C4-N3-C2} & 110.66 & 110.82 & 110.8\\
{C2-N1} & 1.362 & 1.354 & 1.338  & {N3-C2-N1} & 129.26 & 129.21 & 129.0\\
{N1-C6} & 1.360 & 1.355 & 1.349 & {C2-N1-C6} & 117.92 & 118.07 & 118.8\\
{C6-N6} & 1.368 & 1.371 & 1.337 & {C5-C6-N6} & 121.14 & 122.33 & 123.4 \\
{C6-C5} & 1.431 & 1.418 & 1.409  & {N1-C6-C5} & 119.51 & 118.96 & 117.6\\
{C4-C5} & 1.420 & 1.409 & 1.382 & {C4-C5-N7} & 112.28 & 111.52 & 110.7 \\
{C8-H8} & 1.098 & 1.093  & { } & {N7-C8-H8} & 125.17 & 125.00\\
{C2-H2} & 1.108 & 1.098 & { } & {N3-C2-H2} & 115.52 & 115.58\\
{N9-H9} & 1.017 & 1.022 & { } & {C8-N9-H9} & 127.31 & 127.64\\
{N6-H61} & 1.020 & 1.020 & { } & {C6-N6-H61} & 118.87 & 116.04\\
{N6-H62} & 1.021 & 1.020 & { } & {C6-N6-H62} & 119.52 & 117.56\\
\end{tabular}
$^{\rm a}$ This work: DFT geometries obtained with SIESTA, using the PBE~\cite{Perdew} 
functional; 
Ref.~\onlinecite{Santamaria}: geometries obtained by 
Santamar{\'{\i}}a and Vazquez 
using the VWN~\cite{vwn} functional with 
BP~\cite{bp} non-local corrections; 
Exp.: experimental values from crystallized DNA (refs.~\onlinecite{Taylor}
and \onlinecite{Ozeki}).
\end{table}

  
\begin{table}
\caption{Bond distances and angles for isolated guanine$^{\rm a}$.}
\label{gua}
\begin{tabular}{cccccccc}
Distances (\AA) & This work & Ref.~\onlinecite{Santamaria} &
Exp. & Angles (deg) & This work & Ref.~\onlinecite{Santamaria} & Exp.\\
\tableline
{C2-N1} & 1.382 & 1.379 & 1.375  & {C2-N1-C6} & 126.77 & 126.68 & 124.9 \\
{N1-C6} & 1.449 & 1.448 & 1.393 & {N1-C6-C5} & 109.40 & 109.50 & 111.7\\
{C4-N3} & 1.371 & 1.366 & 1.355 & {C4-N3-C2} & 111.66 & 112.24 & 111.8\\
{N3-C2} & 1.335 & 1.324 & 1.327 & {N3-C2-N1} & 124.17 & 123.52 & 124.0\\
{C2-N2} & 1.378 & 1.391 & 1.341 & {N1-C2-N2} & 116.83 & 117.2& 116.3\\
{C4-N9} & 1.386 & 1.380 & 1.377 & {C4-N9-C8} & 106.72 & 106.73 & 106.0\\
{N9-C8} & 1.398 & 1.392 & 1.374 & {N9-C8-N7} & 113.20 & 112.85 & 113.5 \\
{C8-N7} & 1.329 & 1.321 & 1.304  & {C8-N7-C5} & 104.19 & 104.49 & 104.2 \\
{N7-C5} & 1.391 & 1.389 & 1.389 & {N7-C5-C4} & 111.31 & 111.01 & 110.8 \\
{C6-C5} & 1.460 & 1.446 & 1.415 & {C6-C5-C4} & 118.39 & 118.53 & 119.1 \\
{C5-C4} & 1.423 & 1.407 & 1.377 &  {C5-C4-N3} & 129.61 & 129.51 & 128.4 \\
{C6-O6} & 1.237 & 1.234  & 1.239 & {C5-C6-O6} & 131.48 & 131.46 & 128.3\\
{N1-H1} & 1.025 & 1.025 & { } & {C2-N1-H1} & 120.11 & 120.06\\
{N2-H21} & 1.016 & 1.023 & { } & {C2-N2-H22} & 117.27 & 112.9\\
{N2-H22} & 1.017  & 1.023 & { } & {C2-N2-H21} & 122.36 & 116.37 \\ 
{C8-H8} & 1.098 & 1.092 & { } & {N7-C8-H8} & 124.76 & 125.32\\
{N9-H9} & 1.025 & 1.022 & { } & {C8-N9-H9} & 127.56 & 128.11\\
\end{tabular}
$^{\rm a}$ Same as in Table~\ref{ade}.
\end{table}

\hfill\break
\clearpage

\begin{table}
\caption{Bond distances and angles for isolated cytosine$^{\rm a}$.}
\label{cyt}
\begin{tabular}{cccccccc}
Distances (\AA) & This work & Ref.~\onlinecite{Santamaria} & 
Exp. & Angles (deg) & This work & Ref.~\onlinecite{Santamaria} & Exp.\\
\tableline
{N3-C2} & {1.386}& {1.379} & {1.356} & {N3-C2-N1} & {116.77} & {116.47} & {118.9
}\\
{C2-N1} & {1.444} & {1.439} & {1.399} & {C2-N1-C6} & {123.38} & 123.38 & 120.6\\
{N1-C6} & {1.366} & {1.363} & {1.364} & {N1-C6-C5} & {119.91}& 119.71 & 121.0\\
{C4-N3} & {1.341} & {1.332} & {1.334} & {C4-N3-C2} & {119.73} & 119.76 & 120.0\\
{C4-N4} & {1.374} & {1.378} & {1.337} & {N4-C4-N3} & {116.72} & 116.57 & 117.9\\
{C6-C5} & {1.385} & {1.371} & {1.337} & {C6-C5-C4} & {115.87} & 116.18 & 117.6\\
{C5-C4} & {1.456} & {1.445} & {1.426} & {C5-C4-N3} & {124.33} & 124.43 & 121.8\\
{C2-O2} & {1.240} & {1.236} & {1.237} & {N3-C2-O2} & {125.52} & 125.68 & 121.9\\
{N4-H41} & {1.020} & 1.019 & { } & {H41-N4-H42} & {120.31} & 116.37 \\
{N4-H42} & {1.023} & 1.022 & { } & {C4-N4-H42} & {118.29} & 114.91\\
{N1-H1} & {1.027} & {1.023} & { } & {C2-N1-H1} & 115.13 & 115.23\\
{C5-H5} & {1.100} & {1.094} & { } & {C4-C5-H5} & 122.99 & 122.62\\
{C6-H6} & {1.101} & {1.096} & { } & {N1-C6-H6} & 117.23 & 117.29\\
\end{tabular}
$^a$ Same as in Table~\ref{ade}.
\end{table}


\begin{table}
\caption{Bond distances and angles for isolated thymine$^{\rm a}$.}
\label{thy}
\begin{tabular}{cccccccc}
Distances (\AA) & This work  & Ref.~\onlinecite{Santamaria} 
& Exp. & Angles (deg)  & This work & Ref.~\onlinecite{Santamaria} & Exp.\\
\tableline
{C4-N3} & 1.420 & 1.417 & 1.413 & {C4-N3-C2} & 128.46 & 128.17 & 126 \\
{N3-C2} & 1.399 & 1.393 & 1.345 & {N3-C2-N1} & 112.57 & 112.57 & 118 \\
{C2-N1} & 1.407 & 1.398 & 1.314 & {C2-N1-C6} & 123.77 & 123.70 & 123 \\
{N1-C6} & 1.386 & 1.387 & 1.408 &  {N1-C6-C5} & 122.72 & 122.71 & 120 \\
{C6-C5} & 1.379 & 1.364 & 1.369 & {C6-C5-C4} & 118.10 & 118.12 & 119 \\
{C5-C4} & 1.481 & 1.470 & 1.476 & {C5-C4-N3} & 114.37 & 114.69 & 114 \\
{C5-CM} & 1.510 & 1.506  & 1.522 & {C4-C5-CM} & 118.03 & 117.97 & 119 \\
{C2-O2} & 1.235 & 1.233 & 1.246 & {N1-C2-O2} & 123.19 & 123.14 & 122  \\
{C4-O4} & 1.241 & 1.238 & 1.193  & {O4-C4-N3} & 120.35 & 120.07 & 121 \\
{N3-H3} & 1.028 & 1.025 &  & {C2-N3-H3} & 115.46 & 115.59 \\
{N1-H1} & 1.024 & 1.022 & { }  & {C6-N1-H1} & 120.97 & 121.17  \\
{C6-H6} & 1.097 & 1.096 &  & {C5-C6-H6} & 122.17 & 122.16   \\
{CM-HM1} & 1.108 & 1.105 & & {C5-CM-HM1} & 110.61 & 111.10 \\
{CM-HM2} & 1.107 & 1.104 & & {C5-CM-HM2} & 110.63 & 110.00 \\
{CM-HM3} & 1.104 & 1.102 & & {C5-CM-HM3}  & 111.29 & 111.29 \\
\end{tabular}
$^{\rm a}$ Same as in Table~\ref{ade}.
\end{table}

\hfill\break
\clearpage
\narrowtext

\begin{table}
\caption{H-bonds distances (in \AA) and angles 
for SIESTA optimization of DNA base-pairs. D-A and D-H
are the donor-acceptor and donor-hydrogen distances, respectively.}
\label{bases-relax}
\begin{tabular}{lcccd}
Pair & Bond  & D-A &  D-H & Angle\\
\tableline
GCWC     & N2(H)$\cdot\cdot\cdot$O2  & 2.872  & 1.036 & 178.10\\
         & N1(H)$\cdot\cdot\cdot$N3  & 2.913  & 1.057 & 175.98\\
         & O6$\cdot\cdot\cdot$(H)N4  & 2.770  & 1.057 & 179.98\\
GG1      & N1(H)$\cdot\cdot\cdot$O6  & 2.755  & 1.057 & 174.67\\
         & O6$\cdot\cdot\cdot$(H)N1  & 2.756  & 1.057 & 174.59\\
GCNEW    & N1(H)$\cdot\cdot\cdot$O2  & 2.763  & 1.052 & 173.10\\
         & O6$\cdot\cdot\cdot$(H)N1  & 2.824  & 1.054 & 178.79\\
CC       & N4(H)$\cdot\cdot\cdot$N3  & 2.872  & 1.057 & 173.41\\
         & N3$\cdot\cdot\cdot$(H)N4  & 2.872  & 1.057 & 173.41\\
GG3      & O6$\cdot\cdot\cdot$(H)N2  & 3.169  & 1.026 & 167.00\\
         & N7$\cdot\cdot\cdot$(H)N1  & 2.864  & 1.043 & 171.91\\
GA1      & N1$\cdot\cdot\cdot$(H)N1  & 3.103  & 1.042 & 179.82\\
         & N6(H)$\cdot\cdot\cdot$O6  & 2.844  & 1.044 & 179.72\\
GT1      & N1(H)$\cdot\cdot\cdot$O4  & 2.797  & 1.048 & 179.40\\
         & O6$\cdot\cdot\cdot$(H)N3  & 2.839  & 1.058 & 175.18\\
GT2      & O2$\cdot\cdot\cdot$(H)N1  & 2.843  & 1.038 & 178.18\\
         & N3(H)$\cdot\cdot\cdot$O6  & 2.874  & 1.064 & 173.51\\
AC1      & N3$\cdot\cdot\cdot$(H)N6  & 3.007  & 1.039 & 173.69\\ 
         & N4(H)$\cdot\cdot\cdot$N1  & 3.046  & 1.044 & 176.89\\
GC1      & N3$\cdot\cdot\cdot$(H)N2  & 2.873  & 1.049 & 178.50\\
         & N4(H)$\cdot\cdot\cdot$N3  & 3.093  & 1.045 & 175.77\\
AC2      & N6(H)$\cdot\cdot\cdot$N3  & 2.957  & 1.041 & 169.11\\
         & N7$\cdot\cdot\cdot$(H)N4  & 2.994  & 1.048 & 177.18\\
GA3      & N7$\cdot\cdot\cdot$(H)N1  & 3.137  & 1.044 & 175.92\\
         & N6(H)$\cdot\cdot\cdot$O6  & 2.806  & 1.042 & 164.83\\
TAH      & N3(H)$\cdot\cdot\cdot$N7  & 2.828  & 1.066 & 175.94\\
         & O4$\cdot\cdot\cdot$(H)N6  & 2.991  & 1.035 & 170.84\\
TARH     & O2$\cdot\cdot\cdot$(H)N6  & 3.041  & 1.027 & 169.67\\
         & N3(H)$\cdot\cdot\cdot$N7  & 2.861  & 1.060 & 176.67\\
TAWC     & N1$\cdot\cdot\cdot$(H)N3  & 2.859  & 1.070 & 179.35\\
         & N6(H)$\cdot\cdot\cdot$O4  & 2.946  & 1.039 & 174.35\\
TARWC    & N6(H)$\cdot\cdot\cdot$O2  & 3.006  & 1.033 & 171.08\\
         & N1$\cdot\cdot\cdot$(H)N3  & 2.890  & 1.061 & 177.89\\
AA1      & N1$\cdot\cdot\cdot$(H)N6  & 3.049  & 1.041 & 177.53\\
         & N6(H)$\cdot\cdot\cdot$N1  & 3.049  & 1.041 & 177.53\\
GA4      & N3$\cdot\cdot\cdot$(H)N6  & 3.088  & 1.034 & 173.88\\
         & N2(H)$\cdot\cdot\cdot$N1  & 2.963  & 1.044 & 179.31\\
TC2      & N3(H)$\cdot\cdot\cdot$N3  & 3.065  & 1.053 & 168.05\\ 
         & O4$\cdot\cdot\cdot$(H)N4  & 2.822  & 1.041 & 176.27\\
TC1      & N4(H)$\cdot\cdot\cdot$O2  & 2.879  & 1.036 & 174.26\\
         & N3$\cdot\cdot\cdot$(H)N3  & 3.149  & 1.044 & 165.17\\
AA2      & N7$\cdot\cdot\cdot$(H)N6  & 3.051  & 1.037 & 176.28\\
         & N6(H)$\cdot\cdot\cdot$N1  & 3.062  & 1.040 & 166.63\\
TT2      & N3(H)$\cdot\cdot\cdot$O4  & 2.872  & 1.046 & 172.49\\
         & O4$\cdot\cdot\cdot$(H)N3  & 2.872  & 1.046 & 172.48\\
TT1      & N3(H)$\cdot\cdot\cdot$O4  & 2.885  & 1.049 & 169.12\\
         & O2$\cdot\cdot\cdot$(H)N3  & 2.876  & 1.050 & 170.11\\
TT3      & N3(H)$\cdot\cdot\cdot$O2  & 2.881  & 1.049 & 168.35\\
         & O2$\cdot\cdot\cdot$(H)N3  & 2.881  & 1.049 & 168.33\\
GA2      & N6(H)$\cdot\cdot\cdot$N3  & 3.146  & 1.027 & 166.86\\
         & N7$\cdot\cdot\cdot$(H)N2  & 3.006  & 1.040 & 173.24\\
GG4      & N3$\cdot\cdot\cdot$(H)N2  & 3.056  & 1.037 & 179.30\\
         & N2(H)$\cdot\cdot\cdot$N3  & 3.059  & 1.035 & 179.72\\
AA3      & N7$\cdot\cdot\cdot$(H)N6  & 3.070  & 1.031 & 159.56\\
         & N6(H)$\cdot\cdot\cdot$N7  & 3.070  & 1.031 & 159.57\\
2aminoAT & N6(H)$\cdot\cdot\cdot$O6  & 2.921  & 1.034 & 176.95\\
         & N3$\cdot\cdot\cdot$(H)N1  & 2.955  & 1.069 & 178.95\\
         & N2(H)$\cdot\cdot\cdot$O4  & 2.998  & 1.027 & 176.27\\
GG2      & O6$\cdot\cdot\cdot$(H)N1  & 2.960  & 1.035 & 175.10\\
         & N7$\cdot\cdot\cdot$(H)N2  & 3.114  & 1.026 & 167.68\\
GC2      & O2$\cdot\cdot\cdot$(H)N1  & 2.917  & 1.040 & 174.15\\
         & N3$\cdot\cdot\cdot$(H)N2  & 3.139  & 1.027 & 177.41\\
\end{tabular}
\end{table}


\begin{table}
\caption{Relaxed Interaction and Stabilization 
Energies (kcal/mol) for the relaxed base-pair structures.}
\label{ener-relax}
\begin{tabular}{lrr}
Base Pair &  $E_{int}$ & $E_t$\\
\tableline
{ GCWC}     & {-32.2}  & { -27.6} \\
{ GG1}      & { -30.1} & { -26.7} \\
{ GCNEW}    & { -26.3} & { -19.6} \\
{ CC}       & { -21.1} & { -18.5} \\
{ GG3}      & { -18.2} & { -17.7} \\
{ GA1}      & { -17.9} & { -16.4} \\
{ GT1}      & { -18.9} & { -17.8} \\
{ GT2}      & { -17.8} & { -16.8} \\
{ AC1}      & { -16.4} & { -13.8} \\
{ GC1}      & { -18.0} & { -16.0} \\
{ AC2}      & { -17.9} & { -15.7} \\
{ GA3}      & { -16.5} & { -16.2} \\
{ TAH}      & { -17.6} & { -15.6} \\
{ TARH}     & { -16.4} & { -14.1} \\
{ TAWC}     & { -16.3} & { -14.2} \\
{ TARWC}    & { -15.1} & { -14.2} \\
{ AA1}      & { -14.2} & { -13.7} \\
{ GA4}      & { -14.2} & { -13.6} \\
{ TC2}      & { -13.8} & { -11.5} \\
{ TC1}      & { -12.2} & { -10.6} \\
{ AA2}      & { -13.7} & { -13.0} \\
{ TT2}      & { -13.1} & { -10.9} \\
{ TT1}      & { -12.9} & { -10.7} \\
{ TT3}      & { -12.6} & { -11.2} \\
{ GA2}      & { -12.9} & { -11.7} \\
{ GG4}      & {  -9.6} & { -8.0}  \\
{ AA3}      & { -11.6} & { -10.7} \\
{ 2aminoAT} & { -19.6} & { -16.8} \\
{ GG2}      & { -13.1} & { -12.4} \\
{ GC2}      & {-12.3}  & { -10.2} \\
\end{tabular}
\end{table}

\hfill\break
\clearpage
\widetext

\begin{table}
\caption{Guanine-cytosine Watson-Crick base-pair$^{\rm a}$.}
\label{gcwc}
\begin{tabular}{ldddd}
 & HF/6-31G** & DFT (B3LYP) & DFT (VWN-BP) & SIESTA \\
\hline
\vspace{-0.3truecm}
\\
d(N2(H) $\cdot\cdot\cdot$ O2) & 3.017  &  2.930 &  2.930 & 2.872 \\
d(N2,H) & 1.001 & -- & 1.035 &  1.035 \\
d(N1(H) $\cdot\cdot\cdot$ N3) & 3.037 & 2.920 & 2.923 &  2.913 \\
d(N1,H) & 1.008 &  -- & 1.051  &  1.056  \\
d(O6 $\cdot\cdot\cdot$ H(N4)) &  2.921 & 2.780 & 2.785  &  2.770 \\
d(H,N4) & 1.007 & -- & 1.055 & 1.057 \\
\vspace{-0.2truecm}
\\
E$_{\rm int}$  &  -25.5 & -29.6 & -27.7  & -32.2 \\
E$_{\rm T}$    &     -- & -26.3 & -22.9  & -27.6 \\
\end{tabular}
\noindent 
$^{\rm a}$ HF/6-31G**: results obtained at the
HF level, with a 6-31G** basis. From ref.~\onlinecite{Sponer96}.

\noindent DFT (B3LYP): DFT results obtained with the Becke3LYP~\cite{b3lyp}
functional. From ref.~\onlinecite{Sponer96}.

\noindent DFT (VWN-BP): DFT results obtained with the 
VWN~\cite{vwn} functional with BP~\cite{bp} non-local 
corrections. From ref.~\onlinecite{Santamaria}.

\noindent SIESTA: Present results.
\end{table}

\begin{table}
\caption{Thymine-adenine Watson-Crick base-pair$^{\rm a}$.}
\label{tawc}
\begin{tabular}{lddd}
 & HF/6-31G** & DFT (VWN-BP) & SIESTA \\
\hline
\vspace{-0.3truecm}
\\
d(N6(H) $\cdot\cdot\cdot$ O4) & 3.086  &  2.955 &  2.946 \\
d(N6,H) & 0.999 & 1.037 &  1.039 \\
d(N1(H) $\cdot\cdot\cdot$ N3) & 2.988 & 2.66 & 2.859 \\
d(N3,H) & 1.013 & 1.067  &  1.070  \\
\vspace{-0.2truecm}
\\
E$_{\rm int}$  &  -12.4 & -13.9  & -16.3  \\
E$_{\rm T}$    &     -- & -11.9  & -14.2 \\
\end{tabular}
\noindent 
$^{\rm a}$ HF/6-31G**: results obtained at the
HF level, with a 6-31G** basis. From ref.~\onlinecite{Sponer96}.

\noindent DFT (VWN-BP): DFT results obtained with the 
VWN~\cite{vwn} functional with BP~\cite{bp} non-local 
corrections. From ref.~\onlinecite{Santamaria}.

\noindent SIESTA: Present results.
\end{table}


\begin{table}
\caption{Cytosine-cytosine base-pair$^{\rm a}$.}
\label{cc}
\begin{tabular}{lddddd}
 & HF/6-31G** & MP2//HF & MP2 & DFT (B3LYP) & SIESTA \\
\hline
\vspace{-0.3truecm}
\\
d(N4(H) $\cdot\cdot\cdot$ N3) & 3.050  & 3.050 & 2.980 & 2.900 & 2.872 \\
\vspace{-0.2truecm}
\\
E$_{\rm int}$  &  -17.3 & -18.8 & -20.5 & -20.4  & -21.1 \\
E$_{\rm T}$    &     -- & -17.5 & -18.7 & -18.1  & -18.5 \\
\end{tabular}
\noindent 
$^{\rm a}$ HF/6-31G**: results obtained at the
HF level, with a 6-31G** basis. From ref.~\onlinecite{Sponer96}.

\noindent MP2//HF: results obtained at the MP2 level,
with the HF geometries. From ref.~\onlinecite{Sponer96}.

\noindent MP2: results obtained at the MP2 level, with
MP2 geometries. From ref.~\onlinecite{Sponer96}

\noindent DFT (B3LYP): DFT results obtained with the Becke3LYP~\cite{b3lyp}
functional. From ref.~\onlinecite{Sponer96}.

\noindent SIESTA: Present results.
\end{table}


\begin{references}
\bibitem{eshift}
For attaining a desired degree of convergence, 
the cutoff radii of the 
different orbitals depend on the 
species, and also on the specific
orbitals themselves. We have proposed a 
systematic way of finding the different
cutoff radii as a function of 
a single parameter that defines the
finite-range approximation: the energy shift~\cite{Daniel}. 
All pseudoatomic orbitals
shift upwards in energy when confined 
to a sphere; defining this
energy shift gives all the required 
radii, guaranteeing a balanced basis.
\end{references}
\end{document}